\documentclass[pra,showpacs,twocolumn,floats,floatfix]{revtex4}

\usepackage{epsfig}
\usepackage{amsmath}

\begin{document}

\title{Theory for the photon statistics of random lasers}

\author{M. Patra}
\altaffiliation[Present address:~]{Laboratory for Computational Engineering, 
Helsinki University of Technology, P.\,O. Box 9400, 02015 HUT, Finland}
\affiliation{Instituut-Lorentz, Universiteit Leiden,
Postbus 9506, 2300 RA Leiden, The Netherlands}

\begin{abstract} 

A theory for the photon statistics of a random laser is presented. Noise is
described by Langevin terms, where fluctuations of both the electromagnetic
field and of the medium are included. The theory is valid for all lasers with
small outcoupling when the laser cavity is large compared to the wavelength of
the radiation. The theory is applied to a chaotic laser cavity with a small
opening. It is known that a large number of modes can be above threshold
simultaneously in such a cavity. It is shown the amount of fluctuations is
increased above the Poissonian value by an amount that depends on the number
of modes above threshold.

\end{abstract}


\maketitle

\section{Introduction}

A random laser is a laser where the necessary feedback is not due to mirrors at
the ends of the laser but due to random scattering inside the
medium~\cite{wiersma:95a,wiersma:97a,beenakker:98b}. It was long argued how to distinguish such a random
laser from a random medium with amplified spontaneous emission (ASE) --- in the
former, the randomness is essential for
providing feedback, whereas in the latter, scattering only increases the
dwell-time in the medium and thus the amplification factor. Two years ago,
the first experimental proof of a random laser was given~\cite{cao:99a}. It
was demonstrated that the lasing action was indeed due to the randomness of the
medium, by measuring the emitted radiation at different points on the surface of
the sample and showing that the peaks in the radiation spectrum were completely
different at different points.

Earlier experiments~\cite{sha:94a,lawandy:94a,zhang:95a} were only able to
prove ASE in random media, frequently referred to as
``laser-like emission''. In a medium with saturation both
laser action and ASE lead 
to a dramatic narrowing of the emitted light profile upon crossing some
threshold so that this criterion does not necessarily signal a laser. Most
``traditional lasers'' are characterized by emitting coherent radiation
above threshold so that considering only the
intensities and forgetting about the
fluctuation properties is insufficient. Recently the first two measurements on
the photon statistics of a random laser
have been published. The group of Papazoglou reports that
the emitted radiation becomes only partially coherent~\cite{zacharakis:00a}
whereas the group of Cao reports that the statistics become completely
Poissonian~\cite{cao:01a}.

The theoretical description of random lasers has in the past focused on the
light intensity inside in the laser. Photons were considered as classical
particles that diffuse or move in some other way repeatedly through
the sample while being amplified. (The literature on this and similar methods is
numerous, some more general, some more focusing
towards a particular system; see e.\,g. Ref.~\onlinecite{wiersma:96a} for
one of the earlier papers.) In this way the intensity of the emitted radiation
can be computed, confirming the observed narrowing of the emission line far
above threshold. No results for the fluctuation properties, however, can be
derived in this way.
Recently, random lasers are also simulated by the finite-difference time domain
method (FDTD)~\cite{taflove:95}. While this method in principle can incorporate
quantum fluctuations on a microscopic level, the computational effort is
prohibitively large, so that at most two-dimensional samples can be treated
(see e.\,g. Ref.~\onlinecite{cao:00a}), and most of its value is for
one-dimensional applications (see e.\,g. Ref.~\onlinecite{jiang:00a}).
Furthermore, only short time series can be computed with acceptable effort so
that the fluctuation properties of the emitted radiation are not accessible. A
different, analytical, approach to noise in random lasers has recently been put
forward by Hackenbroich et al.~\cite{hackenbroich:01a}. Since they do not
include mode competition, their work is only applicable near
threshold.

For a linear medium, i.\,e. a medium where, in contrast to a laser, saturation
effects can be neglected, the statistics of the emitted radiation can be
computed directly, e.\,g. by the method of input-output
relations~\cite{beenakker:98a}. No theory of comparable power exists for
lasers. The theoretical treatment of ``non-trivial''
lasers has in the past focused on the Petermann factor (see
Refs.~\onlinecite{petermann:79a,siegman:89a,siegman:89b} for a definition). It
is a geometry-related factor that describes by how much the excess noise of the
emitted radiation is larger than for a ``simple'' single-mode laser ---
\emph{assuming} that the non-trivial laser behaves the same way as a 
single-mode laser, which is basically equivalent to neglecting mode competition
effects. (It should be stressed that the Petermann factor only gives information
about the radiation far above threshold; it gives no information on
threshold
behavior.) 
Since the Petermann factor is a geometrical factor it can be computed
for a linear medium and then used for the corresponding system filled with a
medium with saturation. The Petermann factor has been derived for arbitrary geometries (see
e.\,g. Ref.~\onlinecite{bardroff:99a,bardroff:00a}) but also random media could
be treated~\cite{patra:00a,frahm:00a,schomerus:00a}. 

There thus is a need for a theory that allows one 
to compute the photon statistics
of the emitted light for ``non-trivial'' lasers, in particular this includes
random lasers. In this paper such a theory based on Langevin terms, also
referred to as Langevin noise sources, is presented. Langevin terms have
successfully been used to describe the radiation properties of linear media
from a microscopic model~\cite{bardroff:00a}. On a higher level, they were used
to describe random linear amplifying media~\cite{mishchenko:99a} where the
Langevin terms included both fluctuations of the electromagnetic field and
sample-to-sample fluctuations of the properties of the random medium. None of
these theories included saturation effects of the medium so that they break
down when the lasing threshold is approached. Apart from saturation effects for
a single mode, a large number of modes can be above threshold
simultaneously~\cite{misirpashaev:98a}, so that mode-competition is important
and cannot be neglected.

This paper is organized as follows: In Sec.~\ref{secmodel} the model for the
photon statistics inside the laser is
described and the model equations are derived. These are then solved in
Secs.~\ref{seclinear} and \ref{secdiscret}. Sec.~\ref{secaus} adds the necessary
modifications to go from the fluctuations inside the laser 
to the fluctuations of the photocurrent emitted by the laser. Until
this point all results are valid for arbitrary lasers, provided that the
outcoupling is weak and the volume of the lasing medium is much larger than the
cube of the wavelength. In Sec.~\ref{secbeispiel} we show how to apply the
formalism developed in this paper to three exemplary systems and demonstrate 
thereby that it can
indeed describe all relevant properties of lasing action. In
Sec.~\ref{secrandom} the random laser is treated, and its photon statistics are
computed. In Sec.~\ref{secexperiments} we try to explain the
experimental results mentioned above. We conclude in Sec.~\ref{secende}.

\section{Model}
\label{secmodel}

\begin{figure}[b]
\epsfig{file=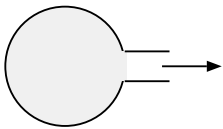,width=3cm}
\caption{\label{figcavity}A (chaotic) cavity is coupled to the outside via 
a small opening. The
cavity is filled with an amplifying medium. The light emitted through the
opening is detected.}
\end{figure}

We consider a optical cavity that is coupled to the outside by an opening
that is small compared to the wavelength of the emitted radiation
(see Fig.~\ref{figcavity}). 
Since the opening is small, there exist well-defined modes in the cavity, 
each with a well-defined eigenfrequency $\omega_i$, $i=1,\ldots,N_{\mathrm{p}}$,
and a eigenmode profile $\Theta_i(\vec{r})$, and all modes are
non-overlapping~\footnote{For cavities with symmetries it can happen that
two modes have the same eigenfrequency but this is not possible for random
cavities
due to mode repulsion~\cite{beenakker:97a}.}.
(In the language of random lasers, this is a ``resonant-feedback
laser''.)
Each mode $i$ thus can be described by the number $n_i$ of photons in it.
Photons in mode $i$ can escape through the opening with rate $g_i$.

\begin{figure}
\epsfig{file=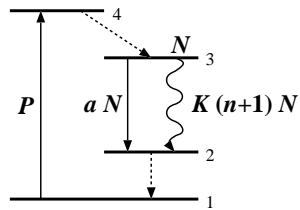,width=3.9cm}
\caption{\label{figmodel}Amplification is modeled by a four-level system, 
where lasing action
(marked by the wiggled line) is from the third to the second level. Dashed lines
mark transitions that are much faster than the other ones and thus need not be
included in the description.}

\end{figure}

The cavity is filled with an amplifying medium.
The medium is modeled by a four-level laser dye (see
Fig.~\ref{figmodel}), where the lasing transition is from the third to the
second level. The transition from the second level to the ground level is 
assumed
to be so fast that the second level is always empty.
The density of excited atoms (i.\,e. atoms in the third level) at
point $\vec{r}$ in the cavity is $N(\vec{r})$. Excitations are created by
pumping with rate $P(\vec{r})$ and can be lost non-radiatively with rate
$a(\vec{r})$.

Coupling between the electromagnetic field and the medium depends on two
quantities, namely the eigenmode profile $\Theta_i(\vec{r})$ of mode $i$, and
the transition matrix element $w(\omega)$ of the atomic transition $3\to2$.
[Frequently $w(\omega)$ will be a 
Lorentzian
centered around some frequency $\Omega$.]
The coupling of mode $i$ to the medium
at point $\vec{r}$ is then given by $K_i(\vec{r}) \equiv w(\omega_i)
|\Theta_i(\vec{r})|^2$. 

The semiclassical equations of motion for $n_i$ and $N(\vec{r})$ are 
(the time
argument for all quantities has been suppressed)
\begin{subequations}
\label{eqklassisch}
\begin{align}
        \dot{n}_i &= -g_i n_i + \int d^3 r \,
                        (n_i+1) K_i(\vec{r}) N(\vec{r}) \;,
                        \\
        \dot{N}(\vec{r}) &= P(\vec{r}) - a(\vec{r}) N(\vec{r})
                - \sum_{i=1}^{N_{\textrm{p}}}
                         (n_i+1) K_{i}(\vec{r}) N(\vec{r}) \;.
\end{align}
\end{subequations}
``Semiclassical'' means that all emission events, pumping events, \ldots are
assumed to be deterministic, with spontaneous emission described by the addition
of a virtual photon to $n_i$ when
computing the transition rates~\footnote{For a cavity with overlapping modes
it is unclear whether the ``$+1$'' would have to be replaced
by ``$+K_i$'' where $K_i$ is the Petermann factor of the $i$-th mode. For
cavities with a small opening this ambiguity does not arise as for them 
$K_i\equiv 1$.}.

To include the randomness of all processes, Langevin terms have to
be added to Eq.~(\ref{eqklassisch}). The four random processes are the escape 
of photons (described by the Langevin term $\Gamma_i$), pumping [described
by $\Phi(\vec{r})$], relaxation of the medium [described by $\alpha(\vec{r})$],
and emission of a photon into mode $i$ at point $\vec{r}$ [described by
$\Psi_{i}(\vec{r})$]. Each of these terms has zero mean, and a correlator that
follows from the assumption that the elementary stochastic processes have
independent Poisson distributions, hence
\begin{subequations}
\label{korrelatorerwartung}
\begin{align}
        \langle \Gamma_i(t) \Gamma_j(t') \rangle &=
                \delta_{i j} \delta(t-t') g_i \langle n_i \rangle
                \label{korrelgamma} \;,\\
        \langle \alpha(\vec{r},t) \alpha(\vec{r}\,',t') \rangle &=
                \delta^3(\vec{r}-\vec{r}\,') \delta(t-t') a(\vec{r}) 
                \langle N(\vec{r}) \rangle
                \label{korrelalpha} \;, \\
        \langle \Phi(\vec{r},t) \Phi(\vec{r}\,',t') \rangle &=
                 \delta^3(\vec{r}-\vec{r}\,') \delta(t-t') \langle P(\vec{r}) 
                 \rangle \label{korrelphi} \;, \\
        \langle \Psi_{i}(\vec{r},t) \Psi_{j}(\vec{r}\,',t') \rangle &=
                \delta_{ij} \delta^3(\vec{r}-\vec{r}\,') \delta(t-t') 
                  \times \nonumber \\
		 &\hspace{1.4cm} K_{i}(\vec{r})\langle (n_i+1) N(\vec{r}) \rangle
                \label{korrelpsi} \;.
\end{align}
\end{subequations}
Eq.~(\ref{korrelpsi}) corresponds with the correlator
given in Eq.~(5b) of Ref.~\onlinecite{mishchenko:99a}.

Adding the terms from Eq.~(\ref{korrelatorerwartung})
to Eq.~(\ref{eqklassisch}) gives the complete equations
of motion:
\begin{subequations}
\label{lasereq}
\begin{align}
        \dot{n}_i &= -g_i n_i + \Gamma_i 
                + \int d^3 \vec{r} (n_i+1) K_{i}(\vec{r}) N(\vec{r})
		\nonumber\\ & \hspace{4.5cm}
                 + \int d^3 \vec{r}  \, \Psi_{i}(\vec{r}) \\
        \dot{N}(\vec{r}) &= P(\vec{r}) + \Phi(\vec{r}) 
                - a(\vec{r}) N(\vec{r}) + \alpha(\vec{r})
		\nonumber\\ & \hspace{1cm}
                - \sum_{i=1}^{N_{\textrm{p}}}
                         (n_i+1) K_{i}(\vec{r}) N(\vec{r}) - 
                \sum_{i=1}^{N_{\textrm{p}}} \Psi_{i}(\vec{r})
\end{align}
\end{subequations}
The sign of the Langevin terms may be chosen freely
as long as the
term $\Psi_i(\vec{r})$ has the opposite sign in the equations for $\dot{n}$ 
and $\dot{N}$.

\section{Linearization}
\label{seclinear}

Eq.~(\ref{lasereq}) cannot be solved by direct numerical methods since Langevin
terms cannot be represented as ``real'' numbers. The only practicable
way to proceed is to linearize the equations. First, we write $n_i=\bar{n}_i +
\delta n_i$ and $N(\vec{r})=\overline{N}(\vec{r}) + \delta N(\vec{r})$, where 
$\bar{n}_i\equiv \langle n_i\rangle$ and $\overline{N}(\vec{r})=\langle
N(\vec{r})\rangle $ are the average solutions. We assume
that these average solutions are identical to the solutions of the 
deterministic rate equation~(\ref{eqklassisch}). This is equivalent to the
factorizing approximation $\langle n_i N(\vec{r})\rangle \approx
 \langle n_i \rangle \langle N(\vec{r})\rangle$.
For a single-mode
cavity like used in cavity-QED this is a bad approximation, leading to errors of
up to a factor $1/4$ in the computed average photon density,
but if the number of modes in the cavity is large
 --- which is the case that we 
are interested in --- this factorization is valid~\cite{rice:94a}.

Inserting this solution, Eq.~(\ref{lasereq}) can be reformulated so that only
$\delta n_i$ and $\delta N(\vec{r})$ remain as variables. Linearization means
that only terms proportional to $\delta n_i$ or $\delta N(\vec{r})$ are kept,
i.\,e. terms proportional to $\delta n_i \delta N(\vec{r})$ are omitted. (This
is justified as long as the variance is sufficiently
smaller than the mean. This condition is equivalent to the
validity of the factorizing approximation used above. It can be
checked self-consistently from the computed results.) This
way one arrives at an equation for the fluctuations alone,
where the coefficients depend on the average solution:
\begin{subequations}
\label{flukeq1}
\begin{align}
        \delta\dot{n}_i &= -g_i \delta n_i + \Gamma_i + \int d^3 \vec{r}       
                (\bar{n}_i+1) K_{i}(\vec{r}) \delta N(\vec{r})
		\nonumber\\ & \hspace{1cm}
                + \int d^3 \vec{r} \delta n_i K_{i}(\vec{r})
                \overline{N}(\vec{r}) + \int d^3 \vec{r} \Psi_{i}(\vec{r}) 
                        \label{flukeq1a} \\
        \delta\dot{N}(\vec{r}) &= \Phi(\vec{r})-a(\vec{r})
                \delta N(\vec{r}) + \alpha(\vec{r})  - \sum_i
                (\bar{n}_i+1) K_{i}(\vec{r}) \delta N(\vec{r}) 
		\nonumber\\ & \hspace{1.7cm}
                - \sum_i \delta n_i K_{i}(\vec{r})
                \overline{N}(\vec{r}) - \sum_{i=1}^{N_{\mathrm{p}}}
                         \Psi_{i}(\vec{r})
                        \label{flukeq1b} 
\end{align}
\end{subequations}

For convenience, we will
label the sum of the Langevin terms in Eq.~(\ref{flukeq1a}) as $f_i$,
and the sum in Eq.~(\ref{flukeq1b}) as $g(\vec{r})$. Evaluating the Langevin terms
from Eq.~(\ref{korrelatorerwartung}) at the average solutions $\bar{n}$ and
$\overline{N}$ gives
\begin{subequations}
\label{expectcorrel}
\begin{align}
        \langle f_i f_j \rangle &=
                \delta_{ij} \left[ g_i \bar{n}_i + \int d^3 \vec{r} (\bar{n}_i
                +1) K_i(\vec{r}) N(\vec{r}) \right] \nonumber \\
                &= 2 \delta_{ij} g_i \bar{n}_i \\
        \langle g(\vec{r}) g(\vec{r}\,') \rangle &=
                \delta^3(\vec{r}-\vec{r}\,') \Bigl[ a(\vec{r})
                P(\vec{r}) + \overline{N}(\vec{r}) 
		\nonumber\\ & \hspace{3cm} + \sum_{i=1}^{N_{\mathrm{p}}}
                         (\bar{n}_i+1)
                K_i(\vec{r}) \overline{N}(\vec{r}) \Bigr]
                \nonumber\\ &= 2 \delta^3(\vec{r}-\vec{r}\,') P(\vec{r}) 
                \nonumber \\ \\
        \langle f_i g(\vec{r}) \rangle &=
                - ( \bar{n}_i + 1 ) K_i(\vec{r}) \overline{N}(\vec{r})
\end{align}
\end{subequations}

\section{Discretization and numerical solution}
\label{secdiscret}

We now discretize the equations in space by picking
points $\vec{r}_j$, $j=1,\ldots,N_{\mathrm{s}}$. Defining $K_{ij}\equiv
K_i(\vec{r}_j)$ and $N_j\equiv N(\vec{r}_j)$ (analogously for all other
quantities), the stationary densities $\bar{n}_i$ and $\overline{N}_j$ from
Eq.~(\ref{eqklassisch}) are the solution of the equations
\begin{subequations}
\label{matrixeq}
\begin{align}
        g_i \bar{n}_i &= \sum_{j=1}^{N_{\mathrm{s}}} (\bar{n}_i + 1 )
                K_{ij} \overline{N}_j \qquad (i=1,\ldots,N_{\mathrm{p}}) \\
        P_j &= a_j \overline{N}_j + \sum_{i=1}^{N_{\mathrm{s}}}
                (\bar{n}_i + 1 ) K_{ij} \overline{N}_j 
                \qquad (j=1,\ldots,N_{\mathrm{s}})
\end{align}
\end{subequations}
This equation cannot be solved analytically but a numerical solution is
straightforward (even though it may be numerically expensive if 
$N_{\mathrm{p}}$ and/or $N_{\mathrm{s}}$ are large).

Eq.~(\ref{flukeq1}) now becomes a linear ODE,
\begin{multline}
        \frac{\mathrm{d}}{\mathrm{d}t}
                \left(\begin{matrix} \delta n_i \\ \delta N_j
                \end{matrix}\right)
        = \\ \left(\begin{array}{cc}
                -g_i + \sum_j K_{ij} \overline{N}_j &
                        (\bar{n}_i+1) K_{ij} \\
                - K_{ij} \overline{N}_j &
                        -a_j -\sum_i (\bar{n}_i+1) K_{ij} 
        \end{array}\right)
        \left(\begin{array}{c} \delta n_i \\ \delta N_j
                \end{array}\right)
        \\ +
        \left(\begin{array}{c} f_i \\ 
                g_j 
                \end{array}\right) 
\end{multline}
where it is understood that all indices $i$ run from $1$ to $N_{\mathrm{p}}$ and
all indices $j$ from $1$ to $N_{\mathrm{s}}$, so that the previous equation can
be written as a $(N_{\mathrm{p}}+N_{\mathrm{s}})$-dimensional matrix equation
$\delta\mathcal{N} = \mathcal{A}\delta\mathcal{N}+\mathcal{L}$. Computing from
$\mathcal{A}$ its matrix $\mathcal{U}$ of eigenvectors and its vector
$\mathcal{E}$ of
eigenvalues, the formal solution can immediately be written down as
\begin{equation}
        \delta\mathcal{N}_j(t) = \sum_{k,l}^{1\ldots N_{\mathrm{p}} + 
        N_{\mathrm{s}}} \int_{-\infty}^{t} \mathrm{d}t' \, \mathcal{U}_{jk}
                \mathrm{e}^{\mathcal{E}_k (t-t')} \mathcal{U}^{-1}_{kl}
                \mathcal{L}_l(t') \;.
        \label{flukdiffgl}
\end{equation}
Since the vector $\mathcal{L}$ consists of Langevin terms, a numerically
computed solution of Eq.~(\ref{flukdiffgl}) is not meaningful. Instead of 
$\delta\mathcal{N}_j(t)$ alone one has to
consider correlators $\langle \delta\mathcal{N}_j(t) \delta\mathcal{N}_{j'}(t)
\rangle$. Noting that the $\mathcal{L}$'s are delta-correlated in time,
and we are interested in $t \to \infty$ (as we are not interested in
intermittent
behavior when switching on the laser) we arrive at
\begin{equation}
        \langle \delta\mathcal{N}_j \delta\mathcal{N}_{j'} \rangle
        = - \sum_{klmn}^{1\ldots N_{\mathrm{p}} + 
                N_{\mathrm{s}}}
                \frac{\mathcal{U}_{jm} \mathcal{U}_{j'n} \mathcal{U}^{-1}_{mk} 
                \mathcal{U}^{-1}_{nl}}{
                \mathcal{E}_m + \mathcal{E}_n }
                \langle \mathcal{L}_k \mathcal{L}_l \rangle \;.
        \label{Ncorrelator}
\end{equation}
Inserting the expectation values of the correlators from
Eq.~(\ref{expectcorrel}) gives the final result where a numerical solution
is easy once the average solution $\bar{n}_i$, $\overline{N}_j$ is
known. ($\langle \mathcal{L}_k \mathcal{L}_l \rangle$ has to evaluated at the
average solution and does thus not depend on time.)


\section{Outcoupling}
\label{secaus}

So far we have considered the number of photons $n_i$ in the $i$-th mode
\emph{inside} the cavity. For practical purposes one is more interested in the
photo current $I$ 
emitted \emph{from}
the cavity. ($I$ gives the number of photons emitted per unit time
and is thus equal to the photon flux integrated over the
entire cross-sectional area.)
Even though the photons from
different modes $i$ are emitted through the same opening, each mode has a
distinct frequency $\omega_i$ so that the modes are easily distinguished on the
outside. We can thus define the photo current $j_i(t)\equiv \bar{j}_i
+\delta j_i(t)$ through the opening due to
the $i$-th mode in the cavity. The photo current can for
example be measured by an (ideal)
photodetector that absorbs the emitted photons.
The fluctuations of the photo current within some time $\tau$ 
(we assume the limit $\tau\to\infty$)
are quantified by the noise power
\begin{equation}
        P_i = \lim_{\tau\to\infty} \frac{1}{\tau} \int_{-\tau/2}^{\tau/2} 
        d t \,
        \delta j(0) \delta j(t) \;.
\end{equation}
The ratio $\mathcal{F}_i = P_i / \bar{j}_i$ is called the Fano factor
and is frequently used to describe the fluctuation properties of optical
radiation. 

\begin{figure}
\epsfig{file=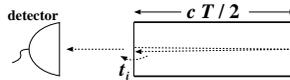,width=3.9cm}
\caption{\label{figneu}The loss rate of photons inside the cavity is given 
by the ratio of
the probability $t_i$ that a photon incident on the outcoupling mirror is
transmitted and the time $T$ needed for one round trip through the cavity.
The photons emitted from the cavity are detected by an ideal photodetector.
}
\end{figure}

In Sec.~\ref{secmodel} we have introduced the loss rates $g_i$. From their
definition it is obvious that the mean 
photo current $\bar{j}_i$ is
\begin{equation}
  \bar{j}_i = g_i \bar{n}_i \;.
\end{equation}
To also compute the fluctuations $\delta j_i$ we need to
treat the outcoupling in more detail.
In a traditional laser (see Fig.~\ref{figneu}) the loss rate $g_i$ is given by 
the ratio of the transmission probability $t_i$ (in classical optics
referred to as ``transmittivity'') 
through the 
outcoupling mirror
and the round-trip time $T$ through the cavity, 
\begin{equation}
        g_i=t_i/T \;.
        \label{outcoupfak}
\end{equation}  
The transmission through the
outcoupling mirror changes the noise of the signal compared to the noise inside
the cavity, and the Fano factor of the emitted radiation is~\cite{patra:00b}
\begin{equation}
        \mathcal{F}_i = t_i
                \frac{\langle \delta n_i \delta n_i \rangle}{\bar{n}_i} 
                + 1 - t_i\;.
        \label{fanoeinzeln}
\end{equation}
This equation can, apart from following the quantum-optical approach of
Ref.~\onlinecite{patra:00b}, also be understood by the following simple
argument: The fraction $\langle \delta n_i \delta n_i \rangle / \bar{n}_i$
on the right-hand side is the Fano factor of the
radiation trapped inside the cavity in mode $i$. With probability $t_i$ the
detector will ``see'' the radiation inside the cavity, and with probability
$1-t_i$ it will see reflected vacuum fluctuations (which have a Fano factor
equal to $1$).

The Fano factor for a measurement where the photons emitted from the cavity in
all modes are detected simultaneously is
\begin{equation}
        \mathcal{F} = \frac{\sum_i t_i^2 \langle \delta n_i \delta n_i \rangle}{
                \sum_i t_i \bar{n}_i } 
                + \frac{\sum_i t_i (1-t_i) \bar{n}_i}{ \sum_i t_i \bar{n}_i }
                \;.
        \label{fanoalles}
\end{equation}

It is immediately obvious that $t_i$ and $g_i$ can for a traditional laser be
identified by properly choosing the unit of time (for the simple laser from
Fig.~\ref{figneu}: by choosing $T$ as the unit of time). We will show in
Sec.~\ref{secrandom} that this is also possible for a random laser. In the
following when giving numerical values or distribution functions
for $g_i$ this identification has been made.

\begin{figure*}[t!]
\epsfig{file=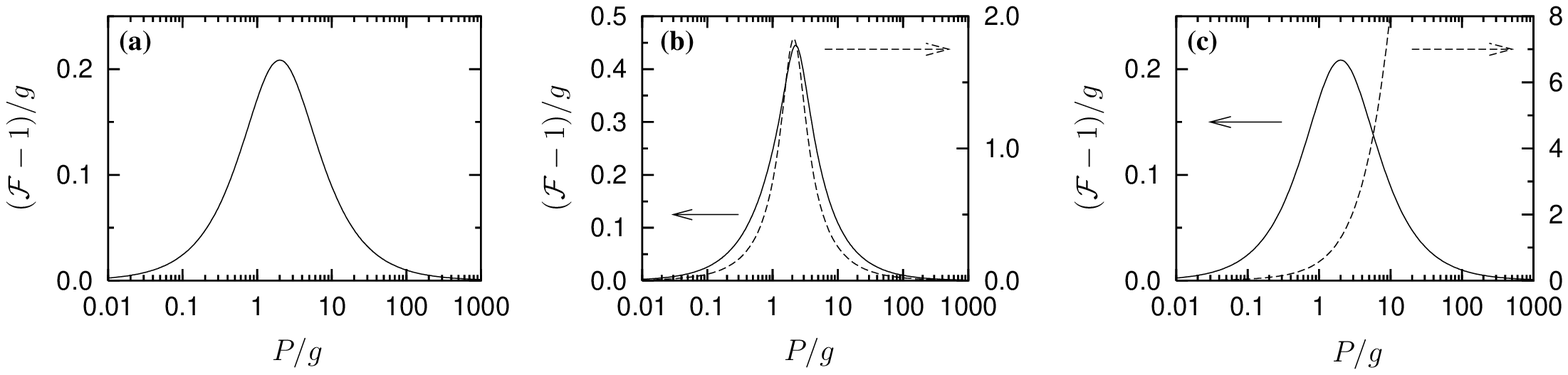,width=6.5in}
\caption{\label{testfaelle}Comparison of the Fano factor $\mathcal{F}$ 
for three different
conditions. The left axis (solid line) depicts the Fano factor for the
integrated emitted radiation, the right axis (dashed line) for the
lasing mode only. \textbf{(a)}~Laser with just a single mode. 
\textbf{(b)}~Laser with a cavity supporting $10$ modes
where one mode is coupled out much less than the others, thus effectively
modeling a
single-mode laser with $\beta\approx 0.1$.
\textbf{(c)}~Laser with $10$ identical modes. }
\end{figure*}

\section{Comparison of lasing regimes}
\label{secbeispiel}

To demonstrate the application of the formalism presented in this paper
and the validity of the
approximations made  in this paper we first want
to discuss three simple cases not involving random media. 
For simplicity we set $a\equiv w\equiv 1$,
$N_{\mathrm{s}}=N_{\mathrm{p}}$ and $K\equiv \mathrm{const}$. This reduces the
number of parameters significantly without reducing the physical content.

The physical features of a laser (in contrast to a linear amplifier) are easily
understood in the following picture: A certain number of excited atoms are created
by pumping within a certain time, and each of those excitations has to be
``consumed'' either by nonradiative relaxation or by emitting one photon from
the cavity. For high photon number in the cavity, nonradiative relaxation can be
neglected, and each pumping event eventually leads to the emission of one photon
from the cavity. The fluctuations of the \emph{integrated} photo current are thus
equal to the fluctuations of the pump source, assumed to be Poissonian throughout
this paper.

In Fig.~\ref{testfaelle}(a) the single-mode laser ($N_{\mathrm{p}}=1$) with a
small opening ($g=10^{-2}$) is treated. The computed curve reproduces the
features of a ``traditional'' laser. The precise location of the maximum is
somewhat off (see the discussion of the factorization approximation above, or
refer to Ref.~\cite{herzog:00a} for a more detailed discussion of the effects
of different approximations on the computed curve near the lasing threshold)
but its height reproduces the exact quantum-mechanical value well. For high
values of the pumping, the photon statistics of the emitted radiation becomes
Poissonian, as qualitatively explained  above.

In Fig.~\ref{testfaelle}(b) we have modeled a laser with one mode coupled to
the outside with $g=10^{-2}$ and the other $9$ modes with $g=10^{-1}$, hence
$N_{\mathrm{p}}=10$. (The value $g=10^{-2}$ was chosen for scaling the axes of
the figure.)
The mode
with the smallest $g$ will be the lasing mode, whereas radiation in the other
modes quickly escapes to the outside so that no significant number of photons
can accumulate in those modes. This basically models a single-mode laser
where only a fraction $\beta=1/N_{\mathrm{p}}$ of the spontaneous radiation is
emitted into the lasing mode. ($\beta$ is called the spontaneous emission
factor. An ideal cavity-QED laser has $\beta=1$ whereas a semiconductor laser
can have a $\beta$ as low as $\beta=10^{-8}$.) The behavior is similar to
Fig.~\ref{testfaelle}(a), except that the peak of the Fano factor of the lasing
mode is larger by about a factor $8$. For small beta, one expects a scaling
$\propto \beta^{-1/2}\approx 3$~\cite{rice:94a} but $\beta=1$ and $\beta=0.1$
are too large for that scaling to be exactly valid.

In Fig.~\ref{testfaelle}(c) the system is kept at $N_{\mathrm{p}}=10$ with
all $g_i\equiv 10^{-2}$. The total radiation depicts the same qualitative behavior
as for the two cases presented so far but the radiation emitted by the lasing
mode alone (in this case: by an arbitrary but fixed mode) depicts a completely
different picture: The Fano factor diverges as the pumping is increased. This
is easily understood by the qualitative description given 
above. For high pumping, every pump
excitation eventually results in one photon being emitted from the cavity, but 
if there are several lasing modes the photon still has the freedom to chose one
of those modes. These additional fluctuations can be that large that they
eventually lead to a very large Fano factor for large pumping. (It is
obvious that the Langevin approach will break down eventually if the
fluctuations become too large, as explained above.)

The three test cases show that the model presented here is able to explain all
relevant features of a laser.

\section{Random laser}
\label{secrandom}

A random laser is a laser where the feedback is not due to mirrors  at the ends
of the laser but due to chaotic scattering, either caused by scatterers placed
at random positions or by a chaotic shape of the
cavity~\cite{wiersma:95a,beenakker:98b}. If the mean outcoupling is weak a
large number of modes in the cavity can be above threshold
simultaneously~\cite{misirpashaev:98a}. As seen above, mode competition
introduces additional noise into the modes. However, even if there are several
modes above threshold, there only will be mode competition \emph{if} the modes
are spatially overlapping and thus are ``eating'' from the same excitations.
The main purpose of this paper is to answer the question whether in a random
laser there is a relevant level of mode-competition noise or whether the
radiation emitted in a laser line approaches Poissonian statistics for strong
pumping -- both statements are mutually exclusively.

We consider a chaotic cavity as depicted in Fig.~\ref{figcavity} 
with a small opening
to the outside. This problem becomes a stochastic problem by considering an
ensemble of cavities with small variations in shape or scatterer positions.
The coefficients appearing in Eq.~(\ref{matrixeq}) thus become random
quantities.
The statistics of these coefficients for a chaotic cavity with small opening is
known~\cite{beenakker:97a,guhr:98a}. The mean loss rate $\bar{g}$ of a cavity
with volume $V$ through a hole of diameter $d$ at frequency $\omega$
is~\cite{bethe:44a}
\begin{equation}
        \bar{g} = \frac{16\pi^2 d^6 \omega^6}{c^6} \cdot
                \frac{\pi^2 c^3}{\omega^2 V^2} \equiv \bar{t} \cdot \delta\;.
        \label{eqgfakt}
\end{equation}
$\delta$ is the level spacing of the cavity. Its inverse
$1/\delta$ is the time needed to explore the entire phase space inside the
cavity and can be identified with the round-trip time introduced for a
``traditional laser''  in Eq.~(\ref{outcoupfak}).

In a chaotic cavity the modes $\Theta_i(\vec{r})$ can be modeled as random
superpositions of plane waves~\cite{berry:77a}. This implies a Gaussian
distribution for $\Theta_i(\vec{r})$ at any point $\vec{r}$~\footnote{If space
is discretized into $N_{\mathrm{s}}$ points $\vec{r}_j$, the distribution of
$\Theta_i(\vec{r}_j)$ is no longer is Gaussian as $\Theta_i(\vec{r}_j)$ now is
the average of $\Theta_i(\vec{r})$ over some region around $\vec{r}_j$. The
most efficient numerical procedure then is to set
$\Theta_i(\vec{r}_1),\ldots,\Theta_i(\vec{r}_{N_{\mathrm{s}}})$ to a random
vector of length $1$, i.\,e. to a column of a random unitary
matrix~\cite{beenakker:97a}.}.
The loss rate $g_i$
is proportional to the square of the gradient of $\Theta(\vec{r})$ 
normal to the opening at the opening, hence its distribution is
\begin{equation}
        \mathcal{F}(g_i) = 
                \frac{\mathrm{e}^{-2 g_i/\bar{g}}}{\sqrt{2\pi g_i \bar{g}}}\;,
        \label{pgeq}
\end{equation}
and $g_i$ and $g_j$ are uncorrelated for $i\ne j$.

It should be noted that the level spacing $\delta$ is no random quantity, so
that $g_i$ and $t_i$ can be identified by choosing $1/\delta$ as the unit of
time.

For simplicity we assume that the amplification profile $w\equiv 1$ so that the
distribution of the
eigenfrequencies is not needed to compute $K_{ij}$. (The distribution is
known~\cite{beenakker:97a} so that an extension to non-constant $w$ is
straightforward.)

\begin{figure}[b]
\epsfig{file=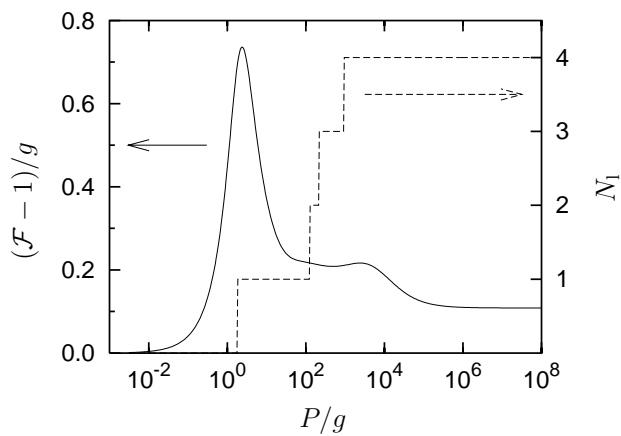,width=3.2in}
\caption{\label{randomsamplefig}Fano factor
of the radiation emitted from the (primary) lasing mode (left axis, solid line)
of some particular sample. The right axis (dashed line) depicts the number of
modes above lasing threshold. Each additional mode crossing the threshold increases the
Fano factor of the primary lasing mode.}
\end{figure}

Fig.~\ref{randomsamplefig} shows the computed Fano factor for a particular
sample from this ensemble ($N_{\mathrm{p}}=10$, $\bar{g}=0.5$ but remember that
the value of $g$ of the lasing mode is much smaller than
$\bar{g}$~\cite{patra:00a,schomerus:00a}). This kind of curve is typical for all
members of the ensemble, while the precise shape varies.  When the first mode
crosses the lasing threshold, the Fano factor goes through a maximum. While
there is a global decrease with increasing pumping, additional peaks are
superimposed each time another mode crosses the lasing threshold. (In the
following a mode is considered to be above lasing threshold if it contains at
least $2$ photons but the results are basically independent of whether one
chooses $1$, $2$ or $10$ photons.) The Fano factor approaches $1$ plus some
finite difference. Mode-competition noise thus gives a contribution to the noise
but there still exists a lasing threshold that is well-defined by a peak of
$\mathcal{F}$.


\begin{figure}
\epsfig{file=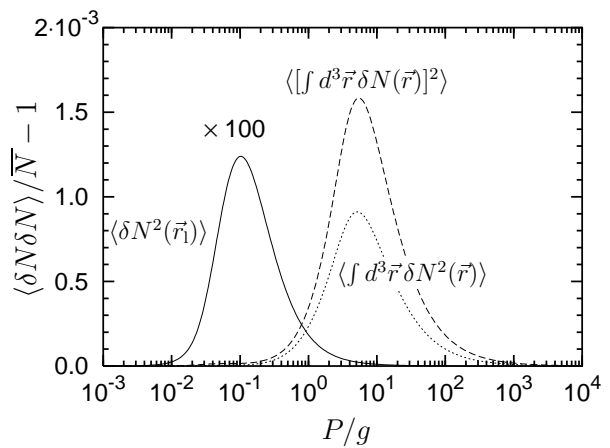,width=3.2in}
\caption{\label{randomsamplemedium}Fluctuations of the excitation density
$N(\vec{r})$ of the medium for the sample from
Fig.~\protect\ref{randomsamplefig}. Depicted are the fluctuations $\langle
\delta N^2(\vec{r}_{\mathrm{l}}) \rangle/\overline{N}(\vec{r_{\mathrm{l}}})$ at
the point $\vec{r}_{\mathrm{l}}$ where the eigenmode profile of the primary
lasing mode has the largest magnitude (solid line, scaled by a factor $100$), 
and the global quantities 
$\langle [ \int d^3\vec{r} \, \delta N(\vec{r}) ]^2 \rangle / \int d^3\vec{r} 
\overline{N}(\vec{r})$ (long dashes) and  
$\langle \int d^3\vec{r} \, \delta N^2(\vec{r}) \rangle / \int d^3\vec{r} 
\overline{N}(\vec{r})$ (short dashes).}
\end{figure}

Similarly to computing the fluctuations of the Fano factor,
it is possible to compute the fluctuations $\delta N(\vec{r})$
of the density of excited atoms directly from Eq.~(\ref{Ncorrelator}). 
Fig.~\ref{randomsamplemedium} depicts the computed fluctuations for the entire 
cavity (dashed lines) as well as for the point $\vec{r}_{\mathrm{l}}$ where the
eigenmode profile $\Theta_l(\vec{r})$ of the primary lasing mode has the 
largest magnitude. The former
quantity peaks at a significantly larger pumping $P$ which is immediately
understood by noticing that the primary lasing mode effects only part of the 
total cavity, and a significant part of the cavity is left 
``untouched'' until more modes have
crossed the lasing threshold.
\begin{figure*}
\epsfig{file=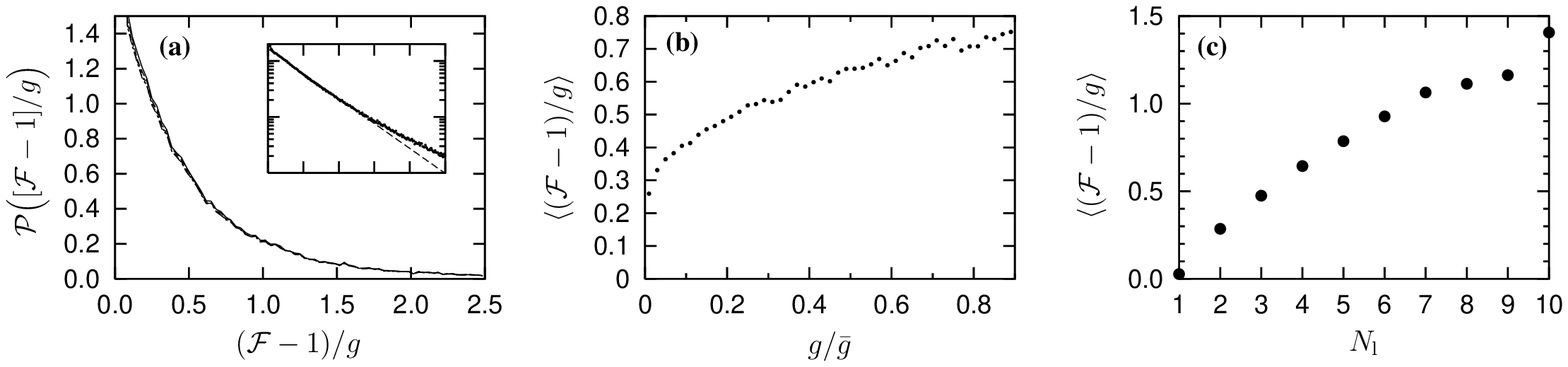,width=6.5in}
\caption{\label{vergleichfig}The value of the Fano factor for the primary 
lasing mode
depends on the number $N_{\mathrm{l}}$ of
cavity modes above laser threshold, not on the other parameters. Unless
otherwise noted, computed from $\approx 9\cdot 10^5$ samples with $\bar{g}=0.1$.
\textbf{(a)}~Probability 
distribution of $(\mathcal{F}-1)/g$ for $\bar{g}=0.1, 0.2,
\ldots, 0.5$. The five curves overlap almost perfectly, 
thereby demonstrating that the size of the opening does not influence the
amount of mode-competition noise generated. (Computed from $\approx 10^5$ samples
for each value of $\bar{g}$ with identical realizations for $K_{ij}$ and 
$g_i/\bar{g}$
for the five runs.)
The inset shows the probability distribution from the large set with
$\bar{g}=0.1$ plotted logarithmically.
\textbf{(b)}~Average of the Fano factor as a function of the
outcoupling constant $g$ of the lasing mode. \textbf{(c)}~Average of the Fano
factor as a function of $N_{\mathrm{l}}$.}
\end{figure*}

The two global quantities depicted, 
$\langle [ \int d^3\vec{r} \, \delta N(\vec{r}) ]^2 \rangle / \int d^3\vec{r} 
\overline{N}(\vec{r})$ and  
$\langle \int d^3\vec{r} \, \delta N^2(\vec{r}) \rangle / \int d^3\vec{r} 
\overline{N}(\vec{r})$, differ by the inclusion of terms $\langle N(\vec{r}_1)
N(\vec{r}_2)\rangle$, $\vec{r}_1 \ne \vec{r}_2$. The different heights of the 
peaks (the first one is higher)
demonstrate that (at least in the relevant interval of $P$, and on average)
the density of excited atoms at different positions is positively
correlated. This can be understood in the following simple picture: The photon
densities $n_i$ and the excitation densities $N(\vec{r})$ are on average
negatively correlated since each emission of an extra photon ($\delta n_i >0$)
leads to the de-excitation of an atom [$\delta N(\vec{r})<0$], and 
vice versa,
hence $\langle n_i \delta N(\vec{r})\rangle <0$. [This has also 
been confirmed by computing
this correlator numerically from Eq.~(\ref{Ncorrelator})]. Since the excited 
atoms
at different positions communicate only via the
radiation field, their density thus has to be positively correlated.

It is difficult to relate the fluctuations of the excitation density of the
medium to the properties of the emitted light. With increasing pumping,
a peak of 
$\langle
\delta N^2(\vec{r}_{\mathrm{l}}) \rangle/\overline{N}(\vec{r_{\mathrm{l}}})$
starts to form (cf. Fig.~\ref{randomsamplemedium}) at the same pumping that a
peak starts to form for the Fano factor $\mathcal{F}$ (cf.
Fig.~\ref{randomsamplefig}) but the location of the maximum of the peak is
significantly different for both curves. The complicated interplay between
radiation modes and matter in a random laser does not allow for a simple
understanding of the relation between these two quantities, and we will not 
discuss the fluctuations of the medium further in
this paper since it focuses on the radiation properties.
The complicated 
structure of the eigenmodes of a chaotic cavity is what makes a random laser 
fundamentally different from a ``traditional'' laser.


In the following we will concentrate on the radiation and on the
Fano factor far above threshold. $P$
is chosen such that $P/g\approx 10^7$ (remember that the value of  $g$ of the
lasing mode fluctuates). This is a compromise between a so large value as possible
to ensure that the limiting value for $P\to\infty$ is approached as
closely as possible, and a not too large value of $P$ to avoid numerical problems
(remember that Fig.~\ref{randomsamplefig} already
spans $11$ orders of magnitude).

The main results of a Monte-Carlo simulation with $N_{\mathrm{p}}=10$ are
depicted in Fig.~\ref{vergleichfig}. 
The scaled Fano factor does not depend on the size of the opening
(Fig.~\ref{vergleichfig}a), and only
weakly on the outcoupling constant of the lasing mode
(Fig.~\ref{vergleichfig}b). As Fig.~\ref{vergleichfig}c clearly shows, the true
dependence is on the number $N_{\mathrm{l}}$ of modes above threshold. (The weak
dependence of the Fano factor on the value of $g$ of the lasing mode can be understood by
noting that $N_{\mathrm{l}}$ is correlated with $g$ of the lasing mode.)
The finite value of $\mathcal{F}-1$ thus indeed
is due to mode competition noise, as
claimed above.

\begin{figure}[b]
\epsfig{file=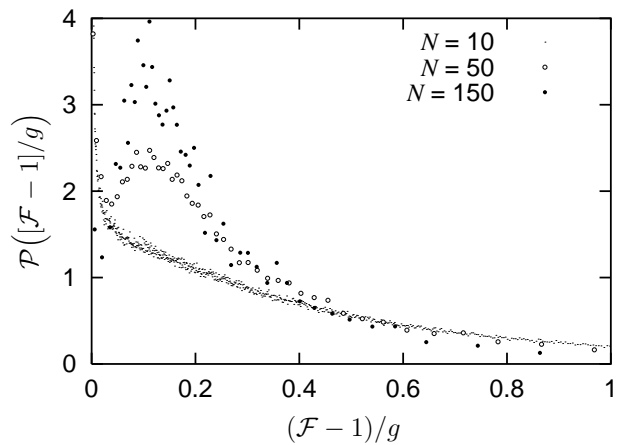,width=3.2in}
\caption{\label{groessenfig}Distribution of the scaled Fano factor for 
cavities with different
number $N_{\mathrm{p}}=N_{\mathrm{s}}$ of modes.}
\end{figure}

For larger cavities, i.\,e. cavities with more modes in it, 
the distribution of $(\mathcal{F}-1)/g$ changes from a peak
near $\mathcal{F}=1$ to one that peaks at a finite value of $(\mathcal{F}-1)/g$,
as seen from Fig.~\ref{groessenfig}. As $N_{\mathrm{p}}$ and $N_{\mathrm{s}}$
increase, the effort to numerically compute the average solution from
Eq.~(\ref{matrixeq}) increases very fast, so that only a comparably small
number of realizations  were computed ($\approx 20000$ for
$N_{\mathrm{p}}=50$ and $\approx 4000$ for $N_{\mathrm{p}}=150$), explaining
the large sampling error in the histograms. [The speed could be increased
significantly by developing an optimized algorithm for solving
 Eq.~(\ref{matrixeq}).] For larger $N_{\mathrm{p}}$ the average of
$(\mathcal{F}-1)/g$ becomes smaller as the large-$\mathcal{F}$ tail gradually
disappears. (From $N_{\mathrm{p}}=10$ to $N_{\mathrm{p}}=150$ the average
becomes smaller by about a factor 2; the average is difficult to compute since
it sensitively depends on few samples with large $\mathcal{F}$.)

\section{Interpretation of experiments}
\label{secexperiments}

\begin{figure}
\epsfig{file=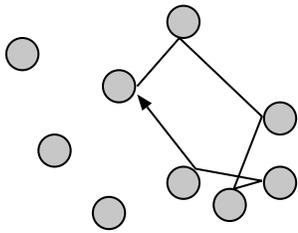,width=4cm}
\caption{\label{virtualfig}Small ``virtual'' cavities can be formed by 
scatterers in the random
medium. Photons can be trapped very efficiently (i.\,e. small
outcoupling) if the distances between the scatterers are compatible with the
wavelength of the radiation).}
\end{figure}

Experiments on random lasers are usually explained by the formation of small
``virtual'' cavities, which can ``trap'' laser light, so that it is scattered
within a small volume many times before it can escape; see 
Fig.~\ref{virtualfig}. (The linear dimension of
such cavities was measured to be of the order of $100$
wavelengths~\cite{cao:99a}). The chaotic cavity used as model in this paper
should be understood as representing one of those virtual cavities. 
It is not obvious which values of the parameters ($N_{\mathrm{p}}$,
$N_{\mathrm{s}}$, $g_i$, \ldots) are needed to explain the experiments.
In the following we will argue that the important parameters are the average
outcoupling $\bar{g}$ and, even more importantly, the probability distribution
$\mathcal{P}(g_i/\bar{g})$ as they together
determine the number $N_{\mathrm{l}}$ of modes above lasing threshold.

Above it was shown that $N_{\mathrm{p}}$ and $N_{\mathrm{s}}$ influence the Fano
factor only weakly, i.\,e. only by a factor $2$, and thus much smaller than the
difference observed in the experiments. Even though it was not explicitly
discussed in this paper, it is obvious that the choice of $w(\omega)$ and
$\Theta_i(\vec{r})$ will not be important, either. This leaves $\bar{g}$ and
$\mathcal{P}(g_i/\bar{g})$ as parameters to explain the experiments.

In this paper, a random laser is modeled by a chaotic cavity with a small
opening. The size of the opening determines the average outcoupling $\bar{g}$,
and all $g_i$ scale linear with $\bar{g}$ [see Eq.~(\ref{pgeq})]. For a virtual
cavity the average outcoupling cannot be computed in such a simple  geometrical
way. The outcoupling $g_i$ for the $i$-th mode in such a virtual cavity depends
delicately on the positions of the scatterers and the wavelength of that mode.
While no theory is available to compute $g_i$ or at least $\bar{g}$
for this case, it is likely that
it will be relatively large as individual scatterers cannot be as effective as
a massive wall with only one small opening.

It was shown in Fig.~\ref{vergleichfig}a that $\mathcal{F}-1\propto \bar{g}$.
This is valid as long as the size of the opening is small compared to the
square of the wavelength. If the opening becomes larger, the modes inside the
cavity acquire a finite width (in frequency space) and start to overlap,
severely complicating the theory~\footnote{Overlapping
modes can exchange particles, so a scattering term would need to be included in
the equations. Even more difficult, the eigenmodes of the cavity no longer are
orthogonal, the Petermann factor thus is larger than
$1$~\cite{patra:00a,frahm:00a,schomerus:00a} and the noise properties change:
More noise is emitted into each mode but the noise in different modes is
correlated so that the total noise power in the linear regime below threshold
stays at the value given by the fluctuation-dissipation theorem. It is not
clear how to include this into the framework presented in this paper.}, and
it is not obvious how the behavior changes. Cao~\cite{cao:01a} speculates that
this overlapping prevents the formation of a fixed photon number
in one particular mode as photons are constantly exchanged between modes with 
nearby
frequencies. Furthermore, the Petermann factor of the lasing mode
becomes significantly large~\cite{frahm:00a} which might or might not increase
the amount of fluctuations. While there is no proof that the amount of
fluctuations is increased by these two effects, it seems to be obvious that
the amount of fluctuations will not decrease due to them. 
Hence, $\mathcal{F}-1$ will at least increase proportional
to the size of the opening --- also for openings that are larger than the
region of validity of the theory presented in this paper.

The previous argument assumes that the number $N_{\mathrm{l}}$ of lasing modes 
inside a virtual cavity is the same as for a chaotic cavity with a small hole.
Mode-overlap itself does not change that number but for a larger opening the
distribution function $\mathcal{P}(g_i/\bar{g})$ 
no longer has the form given by Eq.~(\ref{pgeq}). The
form of $\mathcal{P}(g_i/\bar{g})$ sensitively depends on the kind of 
outcoupling, and
the number of lasing modes in turn sensitively depends on
$\mathcal{P}(g_i/\bar{g})$.
For example, there already is a large difference between a cavity with one
small hole and a cavity with two somewhat smaller holes (so that the total
average loss rate is the same in both cases)~\cite{misirpashaev:98a}. It is very
well possible that the form of $\mathcal{P}(g_i/\bar{g})$ may look
significantly different from Eq.~(\ref{pgeq}) and could depend on many
parameters of the sample.

The differences in $\mathcal{P}(g_i/\bar{g})$ and thus in the number of lasing
modes are thus the natural candidates to explain the differences observed in
the two experiments. 

This prediction could in principle be checked experimentally by measuring the
number of modes above threshold in \emph{one} virtual cavity but to devise an
experimental setup to do this seems very difficult, if at all possible. The
sample used by the group of Papazoglou~\cite{zacharakis:00a} should have
several spatially overlapping modes above lasing threshold (i.\,e. some modes
above threshold are in the same virtual cavity), whereas in the sample by
Cao~\cite{cao:01a} all modes above lasing threshold should be spatially
separated (i.\,e. be in different virtual cavities). One explanation could be
that Cao's sample has more resonant feedback, so that the confinement of
the lasing modes is stronger, compared to Papazoglou's sample. In the latter,
the modes would be extended over a much larger part of the sample (i.\,e. the
virtual cavities are larger), giving them more possibility to overlap.

\section{Conclusions}
\label{secende}

In this paper we have developed a theory to compute the fluctuation properties
of the radiation of a random laser. While for a standard single-mode laser the
emitted radiation becomes coherent far above threshold, the radiation for a
random laser is fluctuating more. It was shown that this extra noise is
due to mode competition noise, i.\,e. due to the uncertainty of deciding into
which mode to photon is emitted by induced emission. 
This noise is larger the higher
the number of modes above lasing threshold is.

To be able to create mode competition noise, the competing modes have to be (at
least partially) overlapping. On the other hand, if the profiles of the modes
are overlapping too much, usually only one of those modes will be above threshold.
The amount of noise created thus is the result of a delicate interplay between
these two competing effects. For a random laser modeled by a chaotic cavity
filled with a laser dye, this leads to a finite increase of the Fano factor far
above threshold, with the precise value depending on the number of modes within
the cavity that are simultaneously above threshold for that particular
realization of the disorder. In particular, the emitted radiation becomes
coherent if only one mode is above threshold.

Recent experiments on random lasers~\cite{zacharakis:00a,cao:01a} gave
conflicting results on whether the noise is increased with respect to the
Poissonian value. Even though it is not directly possible to model the
differences in the two experiments, the theory presented in this paper suggests
that this is due to the differences in the the number of modes above threshold.
This number depends heavily on the specific system in question, so that the
noise properties of a random laser are not universal but depend on the
(experimental) setup.

\begin{acknowledgments}

Valuable discussions with C.\,W.\,J. Beenakker are acknowledged.

\end{acknowledgments}

\addtolength{\textheight}{4mm}

\end{document}